\documentclass{azh}
\usepackage{graphicx}

\begin{document}

\title{Detailed Kinematic Study of the Ionized and Neutral Gas in the Complex of Star
Formation in the Galaxy IC~1613\thanks{
Based on observations collected with  the 6m and 1m telescopes of the Special
Astrophysical Observatory (SAO) of the Russian Academy of Sciences (RAS),
operated under the financial support of the Science Department of Russia
(registration number 01-43)}}

\author{T.A.~Lozinskaya\inst{1}\and A.V. Moiseev\inst{2} \and N.Yu.~Podorvanyuk\inst{1}
}

\institute{
Sternberg Astronomical Institute,
Universitetskii pr. 13, Moscow, 119899 Russia
\and
Special Astrophysical Observatory, Nizhni\u{\i} Arkhyz,
Karachai--Cherkessia, Russia, 369167 Russia
}

\offprints{T.A.~Lozinskaya, \email{lozinsk@sai.msu.ru}}

\date{Received August~20, 2002}

\titlerunning{ Detailed Kinematic  of Gas  in IC 1613}
\authorrunning{ Lozinskaya et al.}

\abstract{
We carried out detailed kinematic studies of the complex of multiple H~I and H~II shells
that represent the only region of ongoing star formation in the dwarf irregular galaxy
IC~1613. We investigated the ionized-gas kinematics by using Fabry--Perot H$\alpha$
observations with the 6-m Special Astrophysical Observatory telescope and the
neutral-gas kinematics by using VLA 21-cm radio observations. We identified three
extended (300\mbox{--}350~pc) neutral shells with which the brightest H~II shells in the
complex of star formation are associated. The neutral-gas kinematics in the complex has
been studied for the first time and the H~I shells were found to expand at a velocity of
15\mbox{--}18~km~s$^{-1}$. We constructed velocity ellipses for all H~II shells in the
complex and refined (increased) the expansion velocities of most of them. The nature of
the interacting ionized and neutral shells is discussed.
}

\maketitle

\section{INTRODUCTION}

The giant complex of multiple ionized shells (Meaburn \emph{et al.}~1988) in the
northeastern sector of IС~1613 is the most prominent structure in narrow-band images of
this Local-Group dwarf irregular galaxy located at a distance of 725\mbox{--}730~kpc
(Freedman~1988a, 1988b; Dolphin \emph{et al.}~2001) in the H$\alpha$, [O~III], and
[S~II] lines.

Most of the bright H~II regions (Sandage~1971; Lequeux \emph{et
al.}~1987; Hodge \emph{et al.}~1990; Price \emph{et al.}~1990;
Hunter \emph{et al.}~1993; Valdez-Gutierrez \emph{et al.}~2001)
and the only known supernova remnant in the galaxy [Lozinskaya
\emph{et al.}~(1998) and references therein] belong to this
complex.

The stellar population of the complex is represented by some twenty young stellar
associations and clusters (Hodge~1978; Georgiev \emph{et al.}~1999; and references
therein).

This multishell complex and the rich stellar grouping represent the only site of ongoing
star formation in the galaxy. This region of violent star formation in IC~1613 can
probably be considered as a very young and small superassociation (Lozinskaya~2002a).

Recent 21-cm radio observations of the complex have shown that extended neutral shells
(supershells in standard terminology) are associated with multiple ionised shells
(Lozinskaya et al 2001, 2002; Lozinskaya 2002b).

The ionized-gas velocities in the region were first measured by
Meaburn \emph{et al.}~(1988); five spectrograms for the bright
part of the complex with poor spatial coverage were used to
determine the characteristic expansion velocities of the H~II
shells, ${\sim30}$~km~s$^{-1}$. Valdez-Gutierrez \emph{et
al.}~(2001) constructed the ionized-gas radial-velocity field in
the H$\alpha$ and [S~II] lines all over the galaxy and estimated
the shell expansion velocities from the splitting of the line
profile integrated over each of the objects.

The H~I and H~II shells close up and partially overlap in the plane of the sky. If their
sizes along the line of sight and in the plane of the sky are assumed to be comparable,
then they can be assumed to be in physical contact with one another\footnote{The galaxy
is inclined at an angle of~$30^{\circ}$ to the plane of the sky and the gaseous-disk
thickness 500\mbox{--}700~pc, as estimated by Afanasiev \emph{et al.}~(2000), is
comparable to the size of the multishell complex. Therefore, a chance projection of
physically unrelated shells located at different distances is unlikely.}.

Deep narrow-band H$\alpha$~images revealed a chain of bright compact emission-line
objects at the boundary of the two closing shells in the complex. Spectroscopic
observations show that the compact objects are early-type giants and supergiants
(Lozinskaya \emph{et al.}~2002). The characteristic ionized- and neutral-gas morphology
suggests that the H~I and H~II shells physically interact in the region of the stellar
chain (see also Section~3).

Thus, the complex of star formation shows evidence of the possible collision between
expanding ionized and neutral shells, the interaction of ionized shells with the
surrounding H~I shells, and the like. This evidence suggests that the birth of
new-generation stars triggered by shell collisions is probable here (Chernin and
Lozinskaya~2002). That is why a detailed study of the structure and kinematics of the
ionized and neutral gas in the multishell complex is of current interest.

The neutral-gas kinematics in the complex of star formation has not yet been
investigated.

Our goal is to study in detail the kinematics of the neutral and ionized gas components
in the shells that constitute the complex of star formation with high spatial and
spectral resolutions.

To study the kinematics of the ionized shells, we carried out H$\alpha$ observations
with a scanning Fabry--Perot interferometer attached to the 6-m Special Astrophysical
Observatory (SAO) telescope. The kinematics of the neutral shells was studied by using
VLA 21-cm radio observations.

The techniques for optical and radio observations and for data reduction are described
in Section~2. In Section~3, we use the ionized- and neutral-gas observations to analyze
the overall structure of the multishell complex in the plane of the sky. The results of
our kinematic study for the H~I and H~II shells in the complex are presented in
Sections~4 and~5, respectively. Our results and conclusions regarding the nature of the
multishell complex that follow from them are discussed in Section~6.

\section{INTERFEROMETRIC H$\alpha$ AND 21~CM OBSERVATIONS}

\subsection*{Fabry--Perot Observations with the 6-m SAO Telescope and Data Reduction}

\begin{table}[t!]
\caption{A log of IFP observations for IC~1613}
\begin{tabular}{c|c|c|c}
\hline
  Date of observation  &IFP type& $T_{exp}$, \AA & Seeing \\
\hline
 Nov.~1, 2000           &  IFP235  & $32\times240$ &   2\farcs0\\
 Sep.~12, 2001           &  IFP501  & $36\times200$ &  1\farcs8\\
\hline
\end{tabular}
\end{table}

Interferometric H$\alpha$ observations were carried out at the
prime focus of the 6-m SAO telescope using a scanning
Fabry--Perot interferometer (IFP). The interferometer was placed
inside the SCORPIO focal reducer, so the equivalent focal ratio
was~($F$/2.9). A brief description of the focal reducer is given
in the Internet
(http://www.sao.ru/~moisav/scor\-pio/scorpio.html); the SCORРIO
capabilities in IFP observations were also described by
Moiseev~(2002). The detector was a TK1024
${1024\times1024}$-pixel CCD~array. The observations were
performed with ${2\times2}$-pixel hardware averaging to reduce
the readout time, so ${512\times512}$-pixel images were obtained
in each spectral channel. The field of view was 4\farcm8 for a
scale of 0\farcs56~per~pixel. An interference filter with
$FWHM=15$~\AA\, centered on the H$\alpha$ line was used for
premonochromatization.

A log of IFP observations is given in Table~1. For our
observations, we used two different Queensgate ET-50
interferometers operating in the 235th and 501st orders of
interference at the H$\alpha$ wavelength (designated in Table~1
as IFP235 and IFP501, respectively). IFP235 provided a spectral
resolution of $FWHM\approx 2.5$~\AA\, near the H$\alpha$ line (or
${\sim110}$~km~s$^{-1}$). The separation between neighboring
orders of interference, $\Delta\lambda=28$~\AA\,, corresponded to
a range of ${\sim1270}$~km~s$^{-1}$ free from order overlapping.
The spectral resolution of IFP501 was ${\sim0.8}$~\AA\, (or
${\sim40}$~km~s$^{-1}$) for a range of $\Delta\lambda=13$~\AA\,
(or ${\sim590}$~km~s$^{-1}$) free from order overlapping.

During the exposure, we sequentially took interferograms of the
object for various IFP plate spacings. Therefore, the number of
spectral channels was 32 and 36 and the size of a single channel
was $\delta\lambda\approx0.87$~\AA\, ($\sim40$~km~s$^{-1}$) and
$\delta\lambda\approx0.36$~\AA\, ($\sim16$~km~s$^{-1}$) for IFP235
and IFP501, respectively.

We reduced our interferometric observations by using the software developed at the SAO
(Moiseev~2002). After the primary data reduction, the subtraction of night-sky lines, and
wavelength calibration, the observational material represents ``data cubes'' in which
each point in the ${512\times512}$-pixel field contains a 32-channel or 36-channel
spectrum. We performed optimal data filtering --- Gaussian smoothing over the spectral
coordinate with $FWHM=1.5$ channels and spatial smoothing by a two-dimensional
Gaussian with $FWHM=2-3$ pixels --- by using the ADHOC software
package\footnote{The ADHOC software package was developed by J.~Boulestex (Marseilles
Observatory) and is publicly available in the Internet.}.

The accuracy of the wavelength calibration using the calibration-lamp line was less than
$3$~km~s$^{-1}$. Our radial-velocity measurements of the night-sky $\lambda6553.617$~\AA
line revealed a systematic shift when measuring the absolute values of the radial
velocities: ${-8\pm3}$~km~s$^{-1}$ for IFP501 and ${15\pm8}$~km~s$^{-1}$ for IFP235. It
should be noted, however, that these values were most likely overestimated, because the
sky-line intensity varied during the scanning time.

The bulk of the observational data used here were obtained with IFP501, which provided a
higher spectral resolution. Only when analyzing the observational data for the supernova
remnant, where the weak line wings were observed at velocities outside the range free
from order overlapping for IFP501, did we additionally use observations with IFP235.

\subsection*{Observations in the 21~cm Line and Data Analysis}

Based on VLA 21-cm observations, we mapped the H~I distribution with a high angular
resolution and studied the neutral-gas kinematics in an extended region of the galaxy
that included the complex of star formation. An application for the project to study the
neutral-gas structure and kinematics in IC~1613 was made by E.~Wilcots; the first
results were published by Lozinskaya \emph{et al.}~(2001).

The data given in Sections~3 and~4 were obtained by combining VLA observations in
configurations~B, C, and~D; the width of a single channel in radial velocity was
2.57~km~s$^{-1}$. We smoothed the data with the Hunning function. The data were
calibrated by the standard method and transformed into maps using the AIPS software
package.

The reduced data are presented in the form of a data cube with an angular resolution of
$7\farcs4 \times7\farcs0$, which corresponds to a linear resolution
of~${\sim23}$~pc.

To map the integrated 21-cm line intensity distribution, we summed only 40 of the
127~possible spectral channels, because no galactic line emission was detected in the
remaining channels.

\begin{figure*}[p!]
\centerline{
\includegraphics[width=10 cm]{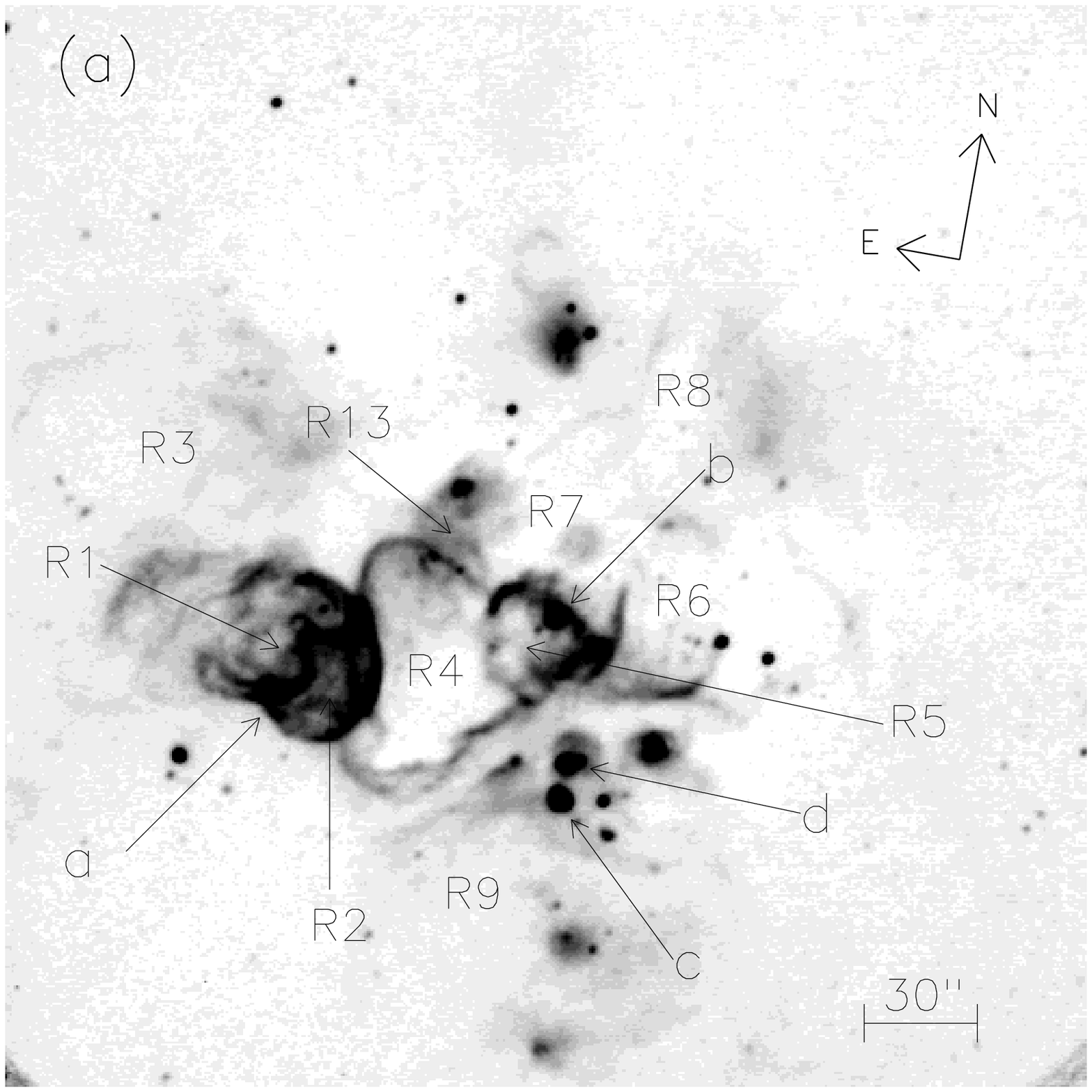}
\includegraphics[width=10 cm]{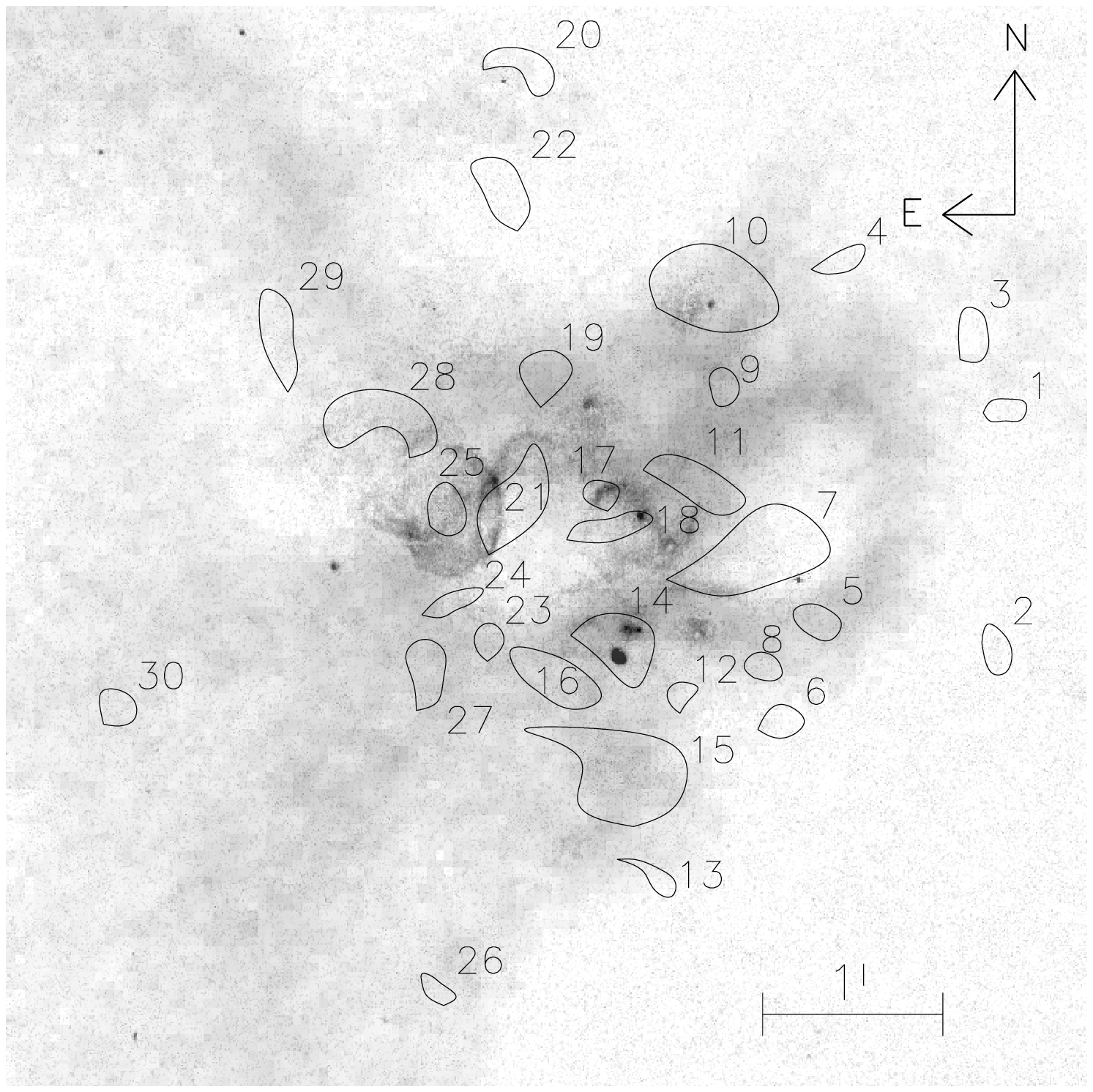}
}
\centerline{
\includegraphics[width=14 cm]{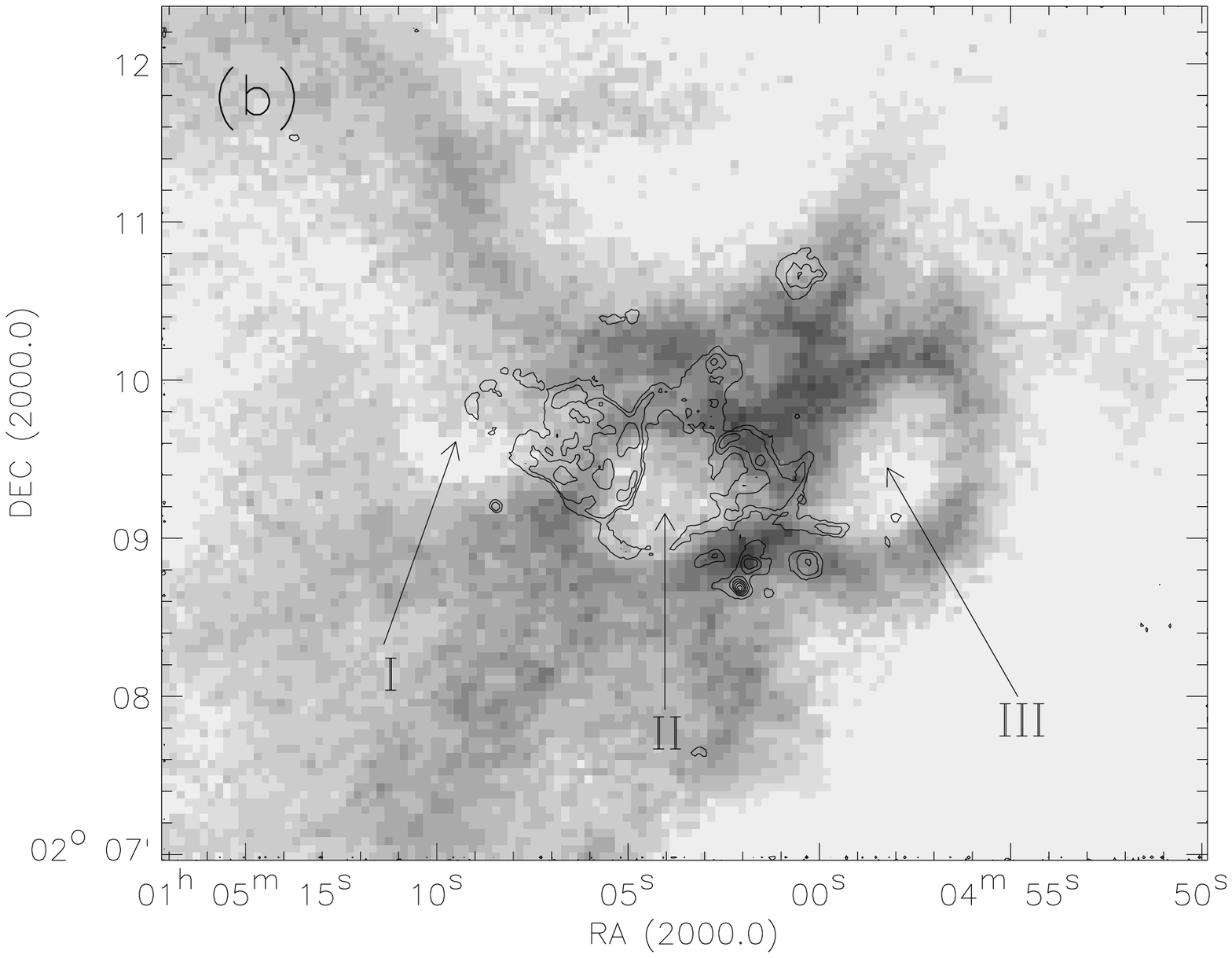}
}
\caption{(a) The monochromatic H$\alpha$ image of the multishell complex obtained with a
Fabry--Perot interferometer on the 6-m SAO telescope. The ionized shells R1\mbox{--}R9
and~R13 from the list of Valdez-Guterrez \emph{et al.}~(2001) studied here are labeled.
The arrows indicate: a "--- the chain of early-type supergiants at the boundary of
shell~R1; b "--- the only Of-type star identified by Lozinskaya \emph{et al.}~(2002); c
"--- the only supernova remnant in IC~1613; d "--- object no.~40a,~b from the list of
H~II regions by Hodge \emph{et al.}~(1990). (b) The H~I intensity distribution
(indicated by shades of gray) superimposed on the monochromatic H$\alpha$ image of the
region (the isophotes corresponding to the brightest regions in Fig.~1a are shown). The
Roman numericals~I, II, and~III mark the three neutral shells investigated here. (c -
{\it top right}) The
boundaries of the OB~associations identified by Georgiev \emph{et al.}~(1999)
superimposed on the intensity distributions in the 21-cm line (indicated by shades of
light gray) and in the H$\alpha$ line (indicated by shades of dark gray) .\hfill}
\end{figure*}

\section{THE OVERALL STRUCTURE OF THE COMPLEX OF STAR FORMATION IN THE PLANE OF THE SKY}

Figure~1а shows the monochromatic H$\alpha$ image of the
multishell complex obtained by integrating the emission over all
spectral channels (in the velocity range from~${-2}$
to~${-584}$~km~s$^{-1}$) using our interferometric observations
with IFP501. The arrows indicate the objects discussed in
Section~5: the chain of early-type giants and supergiants
mentioned above, the only Of~star in the galaxy identified by
Lozinskaya \emph{et al.}~(2002), the only known supernova remnant
--- the nebula~S8 (Sandage~1971), and the bright H~II~region
no.~40a,~b from the list by Hodge \emph{et al.}~(1990). Also
shown in the figure are the ionized-shell numbers from the list by
Valdez-Gutierrez \emph{et al.}~(2001). (Below, we use the
notation adopted in this paper for uniformity.)

All of the brightest ionized shells in IC~1613 are concentrated in the complex of star
formation (Valdez-Gutierrez \emph{et al.}~2001; Lozinskaya \emph{et al.}~2002). In
Fig.~1a, these brightest regions are overexposed to show the faintest H~II shells. The
faint filamentary structures in shells~R6, R7, R8, and R3 are clearly seen in the
figure; we also managed to detect weak emission from thin filaments in several other
regions of the complex.

The chain of bright early-type stars is located at the bright boundary of the ionized
shell~R1 adjacent to shell~R2. This region falls within square~N27 in Fig.~3 from
Valdez-Gutierrez \emph{et al.}~(2001) and includes the nebulae~S10 and~S13, according to
the classification of Sandage~(1971). Association no.~17 from the list by Hodge~(1978)
[its eastern part is designated as no.~25 in the list by Georgiev \emph{et al.}~(1999)]
also lies here.

Our 21-cm H~I observations in IC~1613 with a high angular
resolution allow us to compare the H~I and H~II distributions in
the complex of star formation. The results obtained from these
observations are presented in Figs.~1b and~2a. Figure~1b shows
the 21-cm image of the northwestern galactic sector obtained by
integrating the emission over 40~channels in the velocity range
$-279\div-178$~km~s$^{-1}$ (indicated by shades of gray)
superimposed on the monochromatic H$\alpha$ image (indicated by
isophotes). The latter is represented only by the brightest
regions. The H~I intensity distribution in the entire galaxy is
shown in Fig.~2a (see Section~4).

Even the first observations by Lake and Skillman~(1989) with an angular resolution
of~${60''\times 60''}$ showed that the complex of ionized shells is localized in the
region of the brightest spot on the H~I map of IC~1613. The multishell structure of this
bright spot is clearly seen in Figs.~1b and~2a: the three most prominent H~I shells in
the galaxy surround the chain of bright ionized shells. Below, these bright neutral
shells are called~I, II, and~III for definiteness.

\begin{table*}[t!]
\caption{Parameters of the H~I shells in the complex of star formation}
\begin{tabular}{c|c|c|c|c}
\hline Shell&
 \parbox[c][1cm]{3cm}{central coordinates~(2000)}&
\parbox[c][1cm]{3cm}{Size, arcsec\\ $R$,~pc}&
$V$(exp),~km~s$^{-1}$ &
\parbox[c][1cm]{1.5cm}{Age,\\  Myr}\\
\hline
I       &$1^{\textrm{h}}5^{\textrm{m}}10^{\textrm{s}}$ & $\sim 97$  &               &     \\
        &2$^\circ 9'45''$   & 172~пк     &               &      \\
II      &$1^{\textrm{h}}5^{\textrm{m}}4.5^{\textrm{s}}$&  $73\times81$   &  12--17      &  5.6    \\
        &2$^\circ 9'19''$   &  136~пк              &   &      \\
III     &$1^{\textrm{h}}4^{\textrm{m}}58^{\textrm{s}}$ & $73\times 97$    & 16--18        & 5.3     \\
        &2$^\circ 9'32''$   &  150~пк    &               &      \\
\hline
\end{tabular}

\end{table*}

The central coordinates, sizes, and expansion velocities of the shells estimated in
Section~4 are listed in Table~2. The third column gives the shell sizes along two axes
in arcsec (upper row) and the mean radius in pc determined by them (lower row).

The shells identified in the star-forming region are
300\mbox{--}350~pc in size. These sizes fall within the region of
the peak in the size distribution of giant H~I and H~II shells in
the LMC and SMC (see Meaburn~1980; Kim \emph{et al.}~(1999);
Staveley-Smith \emph{et al.}~(1997).

In addition to these three brightest and most prominent H~I shells, we identified much
larger ring-shaped and arc-shaped structures in IC~1613, with sizes up to
1\mbox{--}1.5~kpc, supergiant shells in the terminology of Meaburn~(1980). These are
clearly seen in the 21-cm image of the entire galaxy shown in Fig.~2a (see also
Section~6).

As follows from Figs.~1a and~1b, a one-to-one correspondence between the neutral and
ionized structures in size and localization is observed only in one case: the ionized
shell~R4 fits well into its surrounding neutral shell~II. The remaining bright H~II
shells are smaller than the neutral shells and are located irregularly: the ionized
shell~R1 lies within the neutral shell~I near its western boundary rather than in the
central cavity; R6 lies within shell~III near its eastern boundary; the ionized
shells~R2 and~R5 fall on the bars between shells~I and~II and between shells~II and~III,
respectively. As was shown by Kim \emph{et al.}~(1999), the giant H~I shells in the LMC
are also interrelated differently with H~II shells and supershells, which reflects the
different evolutionary stages of star formation in the region. This issue is discussed
in more detail in Section~6.

The boundary of the ionized shell~R1 with the chain of blue giants and supergiants
coincides with the thinnest arc-shaped H~I bar between the two H~I shell structures. The
characteristic morphology of the ionized and neutral shells suggests that they are in
physical contact: the thin neutral arc adjoins the ionized shell in the region of the
stellar chain from the outside and the two shells in this region have the same radius of
curvature.

The stellar component of the complex represented by some twenty stellar associations and
clusters is superimposed on the H$\alpha$ and 21~cm images in Fig.~1c. The figure shows
the association boundaries that correspond to a new breakdown of the stellar population
into separate groups in Georgiev \emph{et al.}~(1999). Here, a number of associations
from the list by Hodge~(1978) were split into several smaller groups.

\section{NEUTRAL-GAS KINEMATICS IN THE COMPLEX}

The data cube that we constructed from the 21-cm observations
allows us to analyze the H~I distribution and kinematics in the
entire galaxy. The kinematics of the southeastern region of
IC~1613 around a WO star was studied previously (Lozinskaya
\emph{et al.}~2001). Here, we consider in detail the neutral-gas
kinematics in the complex of star formation for the first time.

Our 21-cm observations show that this complex stands out as the
dynamically most active region in IC~1613. Velocities of internal
neutral-gas motions from $-195\div-200$ to
$-250\div-255$~km~s$^{-1}$ are observed in the region of the
complex ($\Delta RA = 1^{\textrm{h}} 4^{\textrm{m}}
50^{\textrm{s}} - 1^{\textrm{h}} 5^{\textrm{m}} 15^{\textrm{s}},
\Delta D = 2^{\circ} 07' - 2^{\circ} 11'$) against the background
of a smooth systematic variation in the mean H~I velocity along
the galaxy that was pointed out by Lake and Skillman~(1989).

In searching for the possible expansion of the H~I shells in the complex, we constructed
position--radial velocity diagrams in several bands:

1 --- in the band that crosses the pair of shells~I and~II
through their centers;

2 --- in the band that crosses the pair of shells~II and~III in
the same way;

3 --- in the band that passes through the stellar chain at the
boundary of shells~I and~II;

4, 5, and 6 --- in the bands that cross each of the three
shells~I, II, and~III, respectively, in the directions
perpendicular to scans~1~and~2;

7 ---  in the band that includes scans~15 and~16 in the H$\alpha$
line (see Section~5) and that passes through the only supernova
remnant in IC~1613.

The directions of scans~1\mbox{--}7 on the H~I map are shown in Fig.~2a. The
position--radial velocity diagrams constructed for these scans are shown in Fig.~2b.

\begin{figure*}[t!]

\centerline{
\includegraphics[width=10 cm]{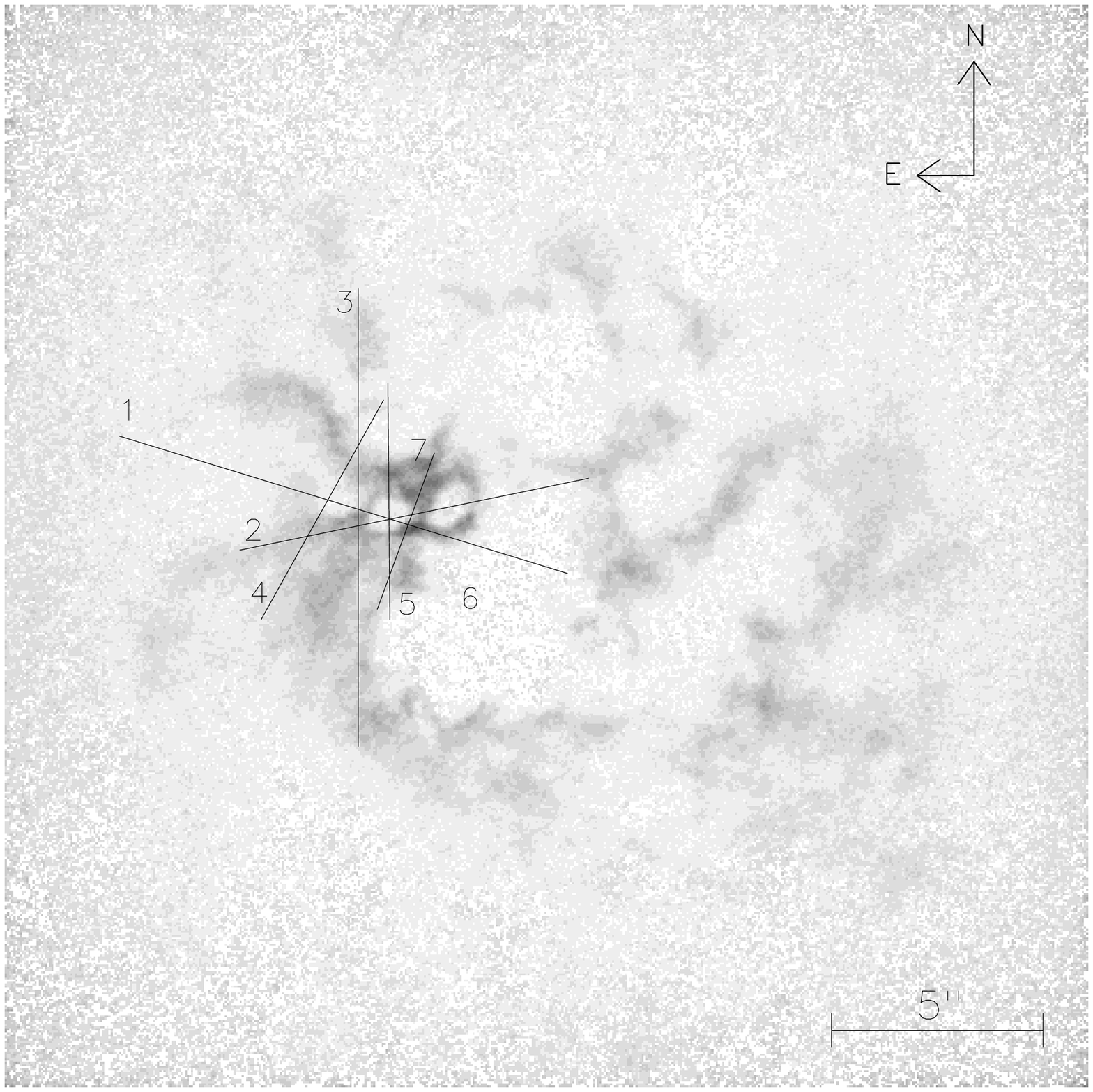}
\includegraphics[width=10 cm]{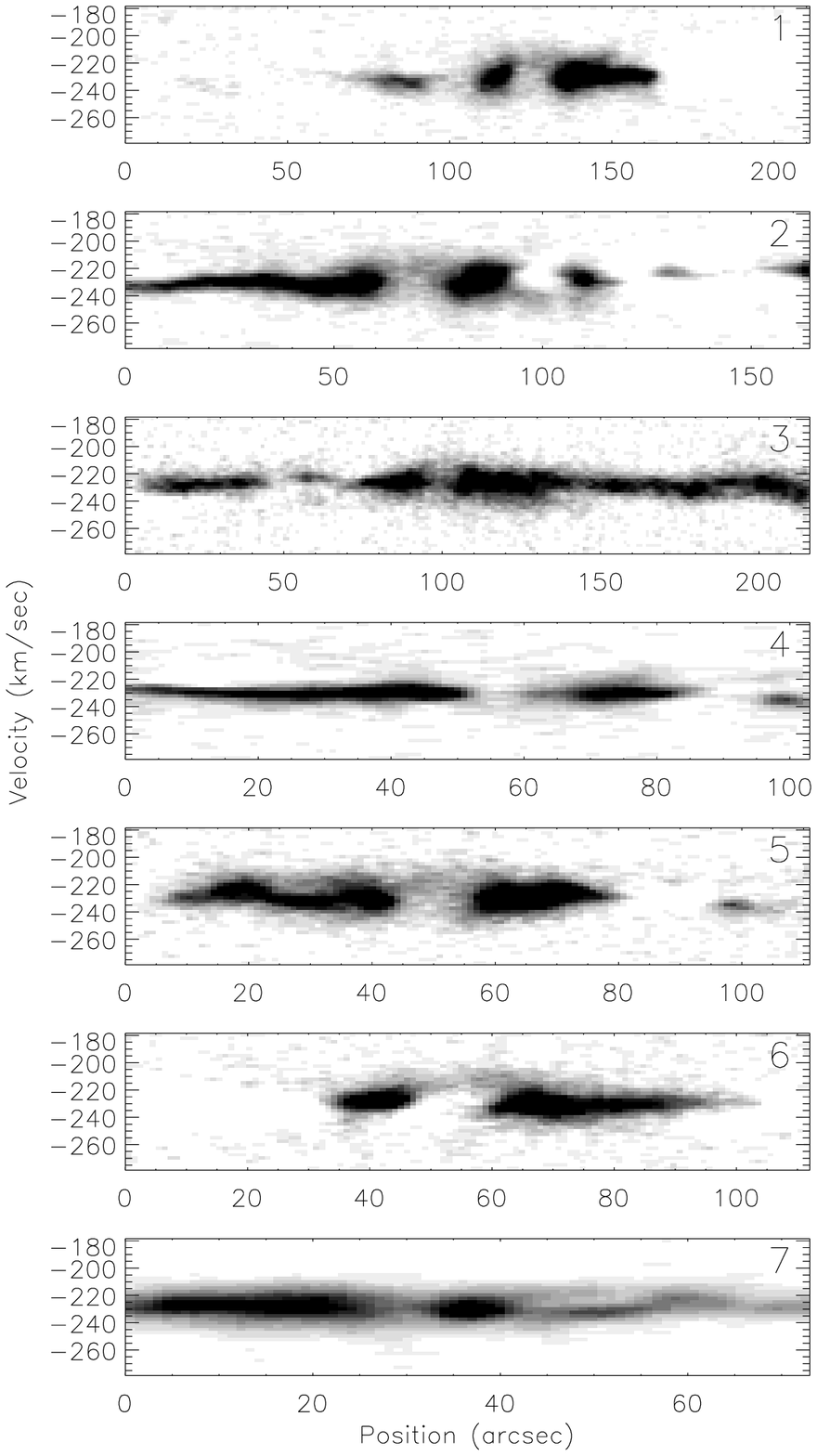}
}
\caption{(a -- {\it left}) The localization of scans~1\mbox{--}7 shown in Fig.~2b on the H~I map of
the galaxy. The numbers are given at the beginning of each scan. (b -- {\it right} ) The position--H~I
velocity diagrams for scans~1\mbox{--}7 (the velocity is heliocentric). \hfill}
\end{figure*}

We chose the band width when scanning in such a way that only the central shell region
and the two peripheral regions in the direction of each scan fell within this width,
which corresponded to 11~pixels or $33''$ in the sky.

The mean H~I velocity in the region of the complex that was determined from the emission
of ``unaccelerated'' gas detected on all scans 1\mbox{--}7 outside shells~I, II, and~III
discussed below is ${V_{\textrm{Hel}} = -230\pm 5}$~km~s$^{-1}$. This value is in good
agreement with the estimate obtained by Lake and Skillman~(1989) for this part of the
galaxy from low-angular-resolution observations.

Figure~2b shows the radial-velocity variation with distance
typical of an expanding shell at least for the two objects
denoted by~II and~III. The characteristic arc-shaped pattern of
velocity variation (half of the ``velocity ellipse'') is seen
both on the scans along the pairs of shells I\mbox{--}II and
II\mbox{--}III and on the scans that cross each of shells~II
and~III in a perpendicular direction (see Fig.~2b). Evidence of
shell-II expansion is seen on scan~1 (370\mbox{--}400~arcsec),
scan~2 (180\mbox{--}230~acrsec), and scan~5
(130\mbox{--}170~arcsec). Scan~2 (280\mbox{--}310~arcsec) and
scan~6 (140\mbox{--}190~arcsec) exhibit shell-III expansion. In
both cases, one side of the expanding shell is clearly seen. The
difference between the unaccelerated-gas velocities on the
periphery of shell~II ($-225\div -230$~km~s$^{-1}$) and on its
approaching side ($-240\div-242$~km~s$^{-1}$) gives an expansion
velocity $V$(exp) ${\sim 12-17}$~km~s$^{-1}$. For shell~III, we
find the expansion velocity from the difference between the
unaccelerated-gas velocities on the periphery ($-228\div
-230$~km~s$^{-1}$) and on the receding side ($-212$~km~s$^{-1}$)
to be $V(exp) \sim 16-18 $~km~s$^{-1}$. The neutral shell~I
exhibits no distinct expansion.

The inferred expansion velocities of the two H~I shells in IC~1613 fall within the
region of the peak in the velocity distribution of neutral supershells in the LMC (Kim
\emph{et al.}~1999).

The only supernova remnant in IC~1613 lies at the outer boundary of the densest H~I
layer in the galaxy that bounds shell~II in the south (see Fig.~2a). Another, fainter
neutral shell in which the supernova remnant is located may adjoin the bright shell~II
in the south. Scan~7 (120\mbox{--}170~arcsec) shows traces of expansion of this fourth
neutral shell. The supernova remnant lies at the inner boundary of this shell
(${\sim120}$~arcsec). This scenario --- a supernova explosion inside a cavity surrounded
by a dense shell and a collision of the expanding remnant with the shell wall
--- was suggested by Lozinskaya \emph{et al.}~(1998) to explain the peculiarity of
this remnant, which combines the properties of young and old objects and which has high
optical and X-ray brightnesses.

The region at the boundary between shells~I and~II where the chain of early-type giants
and supergiants is located exhibits high neutral-gas velocities. A deficit of 21-cm
brightness is observed in this region on scan~3 (280\mbox{--}310~arcsec). This deficit
stems from the fact that emission from only the thin bar between shells~I and~II is
detected here when scanning. However, this weak emission is observed over a wide
velocity range, from ${\simeq -250}$ to ${\simeq -210}$~km~s$^{-1}$ at 20\mbox{--}30$\%$
$I(\max)$.

Our expansion velocities and the corresponding kinematic ages of the bright neutral
shells in the complex of star formation are given in Table~2.

\section{IONIZED-GAS KINEMATICS IN THE COMPLEX}

\begin{figure*}[t!]
\centerline{
\includegraphics[width=10 cm]{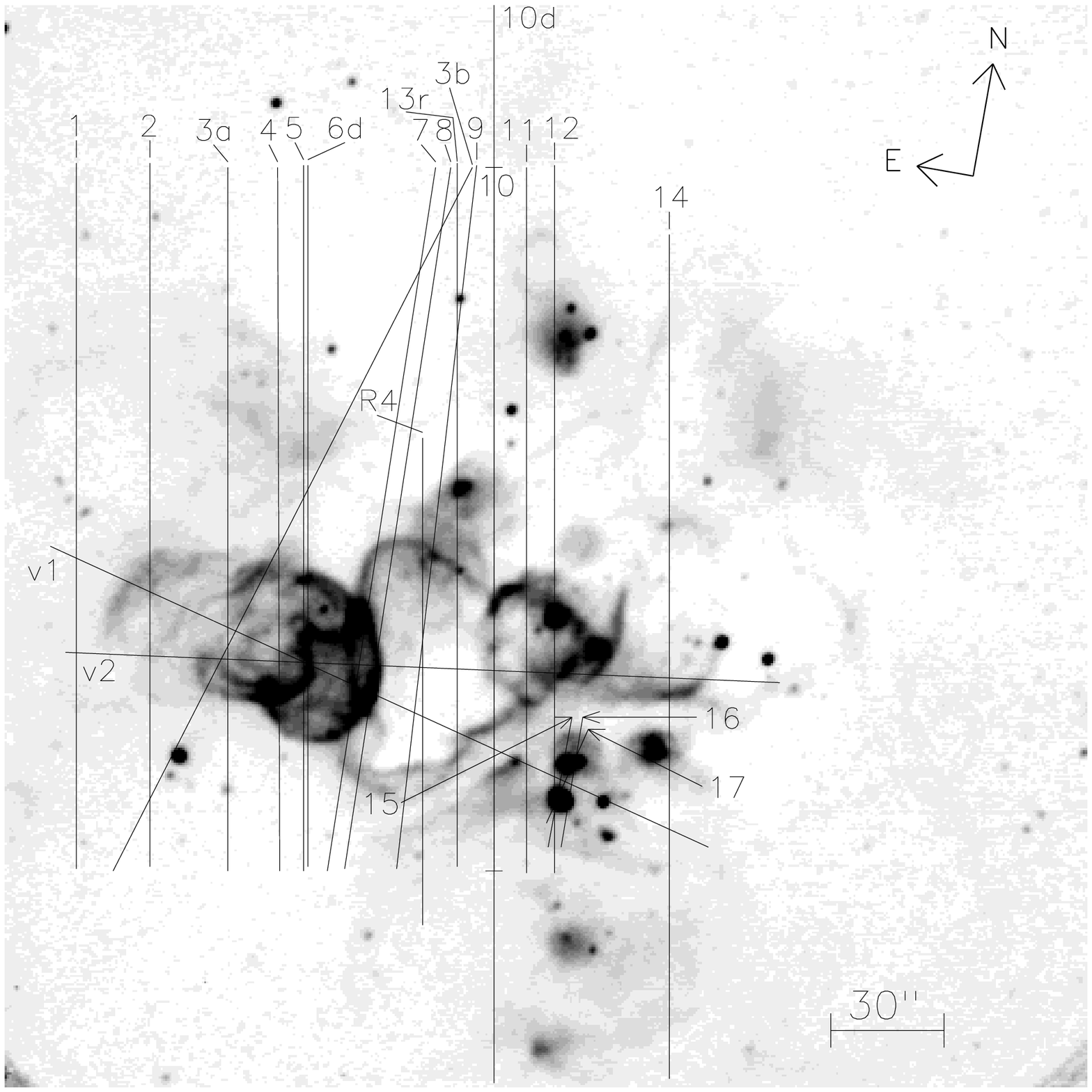}
\includegraphics[width=10 cm]{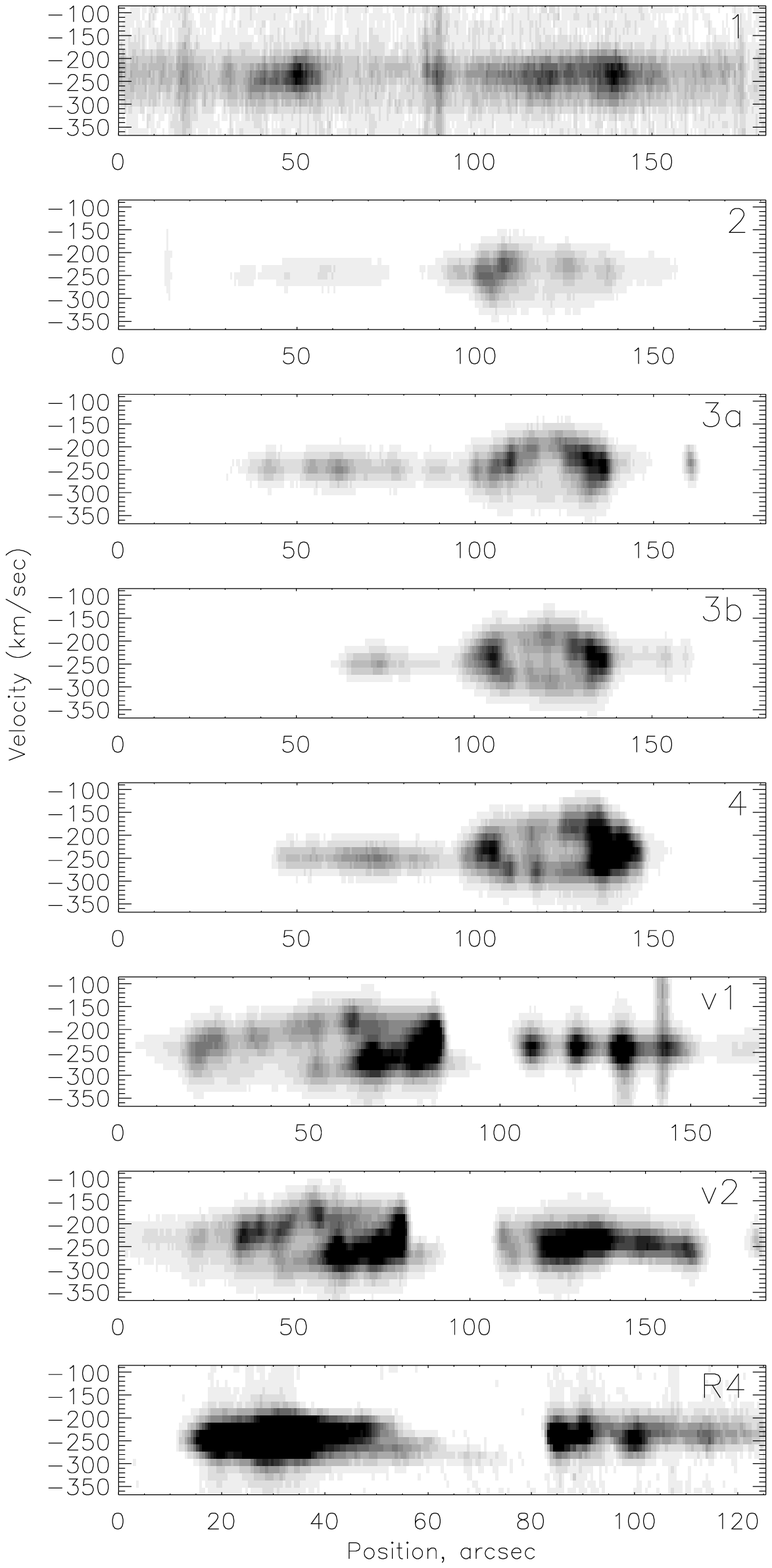}
} \caption{(a -- {\it left})  The localization of the scans shown in Figs.~3b,
3c, and 3d in the H$\alpha$ image of the complex. The numbers are
given at the beginning of each scan. (b -- {\it right}) The position--H~II
velocity diagrams for scans nos.~1, 2, 3a, 3b, 4, v1, v2, and~R4
(the velocity is heliocentric.) \hfill}
\end{figure*}

\addtocounter{figure}{-1}

\begin{figure*}[t!]
\centerline{
\includegraphics[width=10 cm]{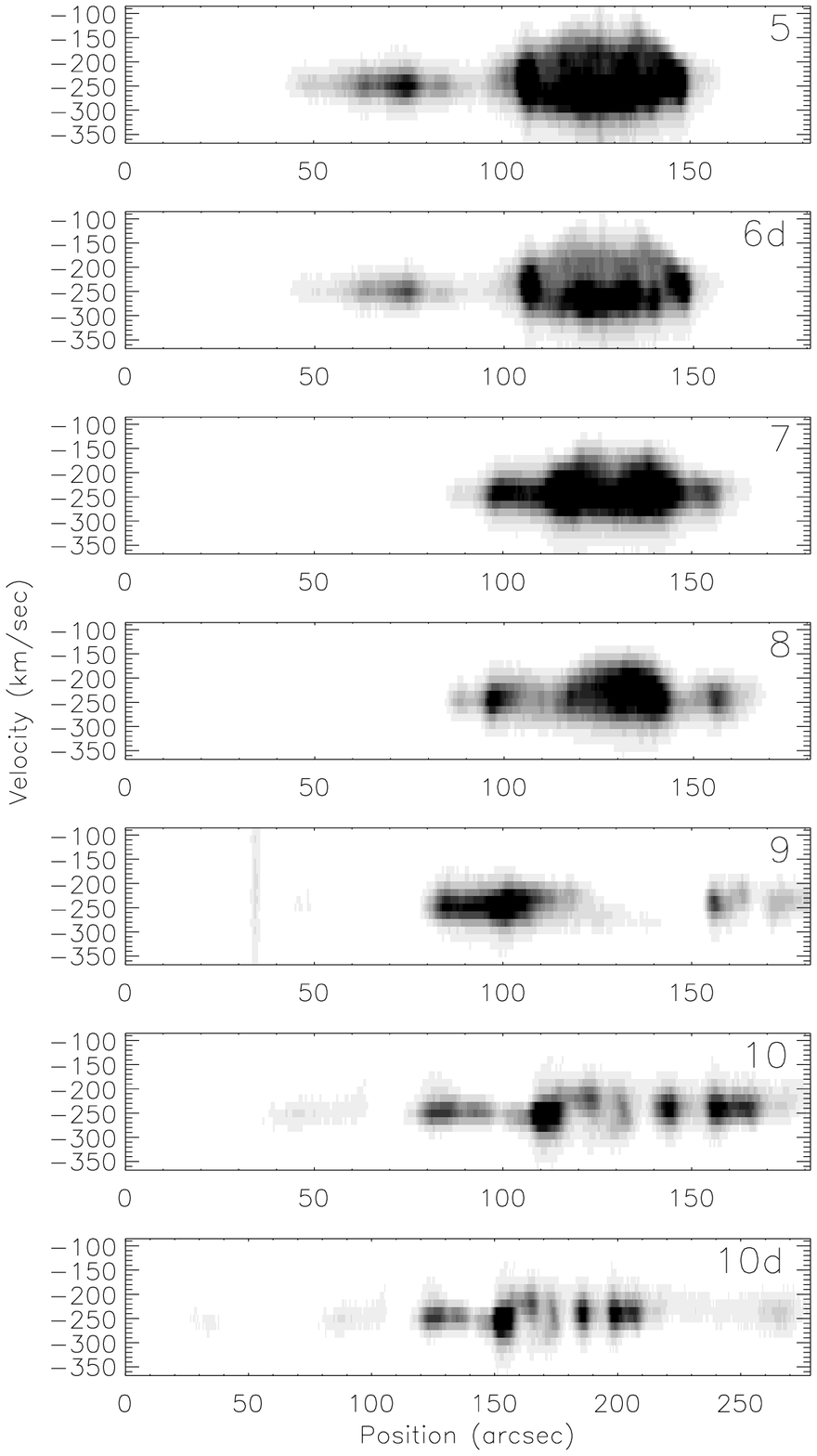}
\includegraphics[width=10 cm]{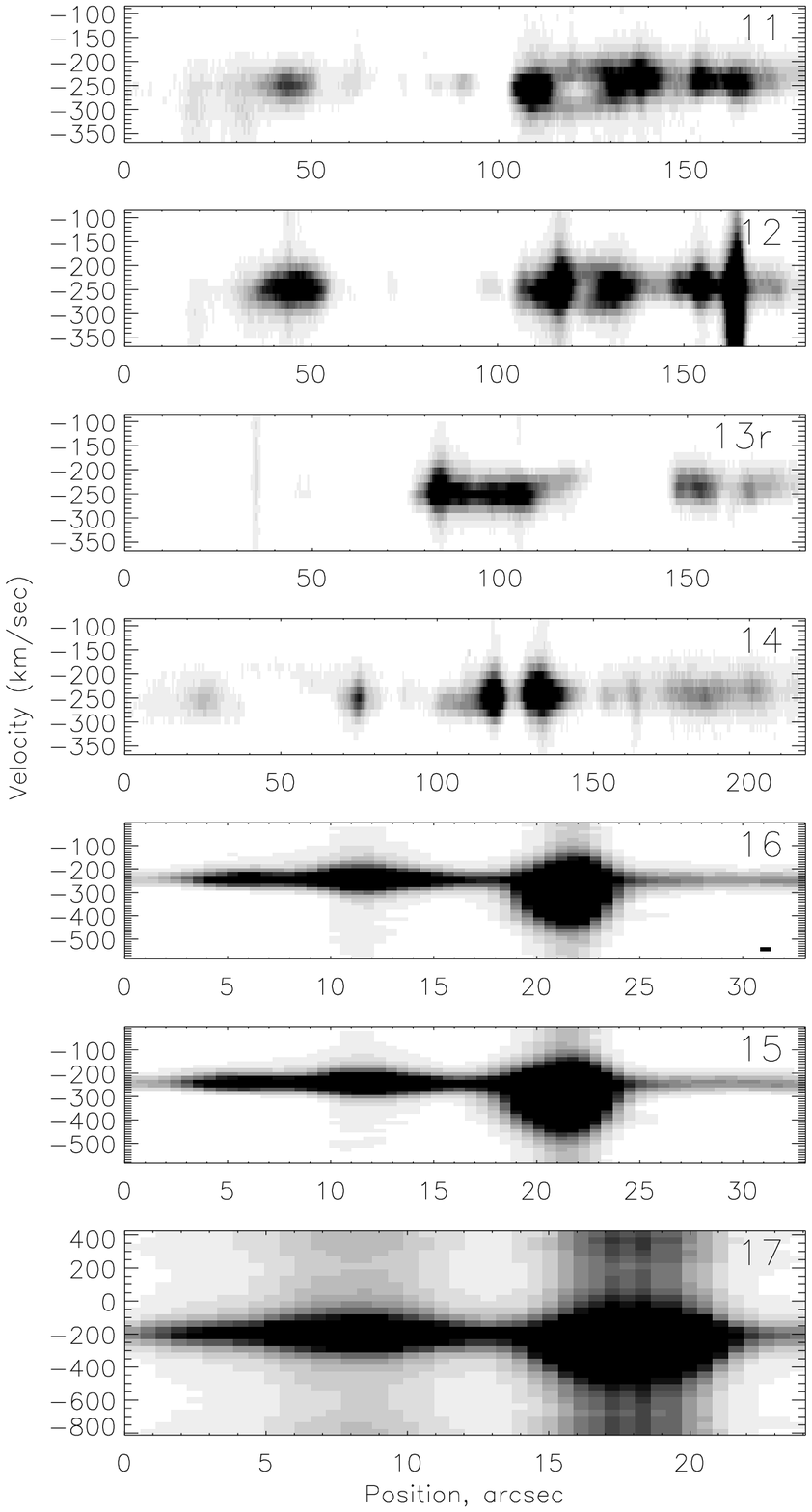}
} \caption{(c -- {\it left}) Same as Fig.~3b for scans nos.~5, 6d, 7, 8, 9, 10,
and 10d. (d -- {\it right}) Same as Fig.~3б for scans nos.~11, 12, 13r, 14, 15,
16, 17. \hfill}
\end{figure*}

To study in detail the kinematics of the ionized shells using our H$\alpha$ observations
with IFP501, we constructed position--radial velocity diagrams for thirty scans, which
cover the entire complex of star formation almost uniformly. For each of the chosen
directions, we constructed two or three scans of different widths: from 1 to 40~pixels
(from $0.5$ to $24''$). Figure~3a shows the localization of some of these scans. The
width of scans~14 and~17 is 21~pixels or $12''$; the width of the remaining scans shown
in Fig.~3 is $11$~pixels or $6''$.

The mean velocity of the unshifted feature in the H$\alpha$
emission  in the complex of star formation is
$-230\div-240$~km~s$^{-1}$. We also observe a smooth velocity
variation in the complex between ${-220}$~km~s$^{-1}$ in the
southeast and ${-260}$~km~s$^{-1}$ in the northwest, in agreement
with the results of Valdez-Gutierrez \emph{et al.}~(2001).

As was shown in Section~3, the elongated system of brightest ionized shells, including~R1,
R2, R4, R5, and~R6, is associated with the neutral shells~I, II, and~III.

Two scans (v1 and v2) cross this chain of bright shells in directions that roughly
coincide with scans~1 and~2 over the neutral shells I\mbox{--}II and II\mbox{--}III,
respectively (see Section~4).

The two scans clearly show H$\alpha$ line splitting along the
entire length of the chain and reveal the characteristic
configuration of the velocity ellipse for individual shells and,
possibly, for the chain of bright shells as a whole. The typical
velocities are about $-290\div-300$~km~s$^{-1}$ for the
approaching sides of the elongated system of bright shells and
about $-190\div-200$~km~s$^{-1}$ for its receding sides. We
successively consider all the ionized shells of this system.

\subsection*{Shells~R1 and~R2 and the Chain of Stars at Their Boundary}

Shells~R1 and~R2 and the region at their boundary where the chain of young stars is
located were scanned repeatedly, in different directions, and in bands of different
widths. All of the scans passing through the central regions of these shells clearly
reveal the characteristic configuration of the velocity ellipse related to their
expansion.

The mean velocity of the unperturbed gas at the eastern boundary
of shell~R1 is ${-240}$~km~s$^{-1}$. The local velocity ellipses
on scan~v2 suggest that the bright western part of shell~R1
(45\mbox{--}55~arcsec) expands at a velocity of
50\mbox{--}60~km~s$^{-1}$ relative to this unshifted feature: the
velocity of the approaching side is ${-290}$~km~s$^{-1}$. In
shell~R2 (55\mbox{--}80~arcsec), the velocities of the bright
approaching and faint receding sides are $-280\div-300$ and
${-170}$~km~s$^{-1}$, respectively. On scan~v1 that crosses the
same region in a different direction, we see a similar arc
structure in region~R2 and individual clumps in region~R1.

Several scans that cross the system of bright shells in the
directions perpendicular to~v1 and~v2 also clearly reveal an
expansion of shells~R1 and~R2. Scans~3a, 3b, and~4 clearly show
that shell~R1 expands at a velocity of 60\mbox{--}70~km~s$^{-1}$
(up to 100~km~s$^{-1}$ in its bright southern part). The bright
regions on the shell periphery exhibit the unshifted line
component at a velocity of $-230\div-250$~km~s$^{-1}$; the
velocities of the two sides of the expanding shell are ${-180}$
and ${-320}$~km~s$^{-1}$ (scan~3). Scan~4 also shows that the
bright southern part of shell~R1 expands faster than does its
northern part: the velocity of the approaching side is the same
in the entire shell, being about ${-300}$~km~s$^{-1}$; the
velocity of the receding side is ${-150}$~km~s$^{-1}$ in the
south and ${-180}$~km~s$^{-1}$ in the north. Scan~2 that crosses
the fainter eastern half of shell~R1 also reveals evidence of its
expansion. The unshifted features on the shell periphery have a
velocity of ${-240}$~km~s$^{-1}$, the bright clump in the
northern part is observed at a velocity of ${-215}$~km~s$^{-1}$,
and the faint approaching side has a velocity of
${-300}$~km~s$^{-1}$.

Valdez-Gutierrez \emph{et al.}~(2001) identified two components in
the integrated line profile at velocities of ${-216}$ and
${-274}$~km~s$^{-1}$ in shell~R1 and two features at velocities
of ${-244}$ and ${-147}$~km~s$^{-1}$ in shell~R2. All these
values fall within the range determined by the velocity ellipses
for these two shells.

The region at the boundary between shells~R1 and~R2 that, in the
21-cm line, corresponds to the thin bar between the neutral
shells~I and~II, where the chain of early-type stars is located,
is represented by scans~5 and~6d. Scan~6d exhibits a bright
emission feature in the H$\alpha$ line at a velocity of about
$-270\div-280$~km~s$^{-1}$, which corresponds to the bright part
of shell~R1 with the chain of stars, traces of stellar continuum
emission, and a weaker emission feature in the region of shell~R2
at higher velocities, up to $-130\div-150$~km~s$^{-1}$. This
entire region on scan~6d (105\mbox{--}150~arcsec) widens toward
the center, with the dense clumps on the periphery and the gas in
the remote part (at 55\mbox{--}85~arcsec) emitting at a velocity
of $-240\div-250$~km~s$^{-1}$. The observed picture suggests an
asymmetric gas expansion in shell~R2 at a velocity of about
30~km~s$^{-1}$ into a denser medium and at a velocity of about
100\mbox{--}110~km~s$^{-1}$ into a less dense medium. We
associate this effect with shell~R2, because the boundary of the
expanding region on scan~6d coincides with the boundary of this
shell. Scan~5 also exhibits irregular motions of the receding
gaseous clumps up to velocities of $-150\div-160$~km~s$^{-1}$;
these are probably related to the local action of the wind from
the chain stars. The velocity of the near side of shell~R2, which
is brighter and more regular, reaches $-280\div-290$~km~s$^{-1}$.
Since the bright clumps on the periphery of the velocity ellipse
coincide with the boundaries of shell~R2, which is larger than
shell~R1 here, we believe that the velocity ellipse on scans~6
and~5 also reflects the expansion of shell~R2.

\subsection*{Shell~R4 and its Possible Collision with Shell~R2}

R4 is the most extended shell in the chain of bright shells and the only one that
completely fits into the surrounding neutral shell. Its central regions are
characterized by weak emission. Scans~R4, 9, v1, v2, and 13r show a deficit of gas in
the central cavity and intense emission on the periphery at a velocity of about
${-240}$~km~s$^{-1}$. The weak H$\alpha$ emission features in the central cavity exhibit
a blueshift and a redshift relative to this velocity. Weak diffuse emission is observed
at velocities up to ${-200}$~km~s$^{-1}$ (scan~R4, 40\mbox{--}55~arcsec; scan~9,
110\mbox{--}120~arcsec) in the north of the cavity and up to ${-290}$~km~s$^{-1}$
(scan~4R, 55\mbox{--}70~arcsec; scan~9, 120\mbox{--}130~arcsec) in the south. The shell
expansion velocity determined by these features is ${\sim45}$~km~s$^{-1}$.

The monochromatic H$\alpha$ image shows a characteristic lenticular structure at the
boundary between shells~R2 and~R4. This structure can be a manifestation of the dense
gaseous ring formed in the region of head-on collision between these shells and oriented
almost edge-on to the observer. The formation of this ring follows from
three-dimensional numerical simulations of the collision between two expanding shells
[see Chernin \emph{et al.}~(1995) and references therein).

In the region that corresponds to the eastern boundary of shell~R4 (scan~7,
95\mbox{--}115~arcsec), we see no significant deviations from the mean ionized-gas
velocity of ${-240}$~km~s$^{-1}$ in the complex. The inclined ring in the region of
collision with shell~R2 is observed as two bright unaccelerated clumps and high-velocity
motions between them (scan~7, 120 and 140~arcsec) and as a bright structure with high
velocities of the internal motions (scan~8, from 120 to 140~arcsec). The velocities of
the faint features between the bright clumps are at a maximum; they reach ${-140}$ and
${-340}$~km~s$^{-1}$ at $7\%$ $I(\max)$ on scan~7a and ${-160}$~km~s$^{-1}$ at $10\%$
$I(\max)$ on scan~8.

\subsection*{Shell~R5}

A characteristic arc structure (110\mbox{--}135~arcsec on scan~10 and
150\mbox{--}175~arcsec on scan~10d) is observed at the boundary between shells~R4
and~R5. This structure may represent the half of the velocity ellipse that corresponds
to the receding side of shell~R5. We estimated the expansion velocity to be
40\mbox{--}50~km~s$^{-1}$.

Scan~11 (105\mbox{--}135~arcsec) also shows a symmetric expansion of shell~R5 at a
velocity of about 35\mbox{--}45~km~s$^{-1}$; the velocities of the two sides of the
shell are ${-220}$ and ${-310}$~km~s$^{-1}$.

A compact group of stars with the only Of~star identified in the galaxy (Lozinskaya
\emph{et al.}~2002) is located on the periphery of shell~R5. Scan~12 shows this star in
a region of 118~arcsec; the star is surrounded by a bright compact H~II region. Given the
IFP501 instrumental profile, the line width in the H~II region at $0.5 I(\max)$ is
${50\pm5}$~km~s$^{-1}$.

\subsection*{Shell~R13 and the H~II Region no.~55 from the List of Hodge
\emph{et al.}~(1990) }

Shell~R13, which adjoins shell~R4 in the north, consists of two
components: a diffuse shell and the bright compact H~II region
nos.~55 (Hodge \emph{et al.}~1990). The bright H~II region shows
a mean velocity of about $-230\div-240$~km~s$^{-1}$ [in agreement
with the measurements of Valdez-Gutierrez \emph{et al.}~(2001)]
and stellar continuum at the center (scan~13r, position
85~arcsec). A group of stars may be located here, because the
area of stellar continuum is broader than that in other regions
of the same scan. The line width in the bright H~II region
reaches 57~km~s$^{-1}$ at $0.5 I(\max)$.

The diffuse shell~R13 also exhibits the mean radial velocity of
$-230\div-240$~km~s$^{-1}$ typical of the entire complex. We
found no clear evidence of shell-R13 expansion.

\subsection*{Shells~R6 and~R8}

According to our data, the expansion velocity of shell~R6 does
not exceed 20~km~s$^{-1}$: the velocities of the northern and
southern shell boundaries (scan~14, 75 and 120~arcsec,
respectively) and the velocity in the central region (about 100
arcsec on the same scan) are about ${-255}$~km~s$^{-1}$.

Shell~R8 exhibits the characteristic arc-shaped pattern of radial-velocity variation of
the far side. The velocity of the bright boundary regions (scan~14, 25 and 75~arcsec) is
${-255}$~km~s$^{-1}$ and the velocity of the faint far side reaches
${-190}$~km~s$^{-1}$, which gives an expansion velocity of about 65~km~s$^{-1}$. We
detected no emission from the even fainter near side.

\subsection*{The Bright H~II Regions nos.~40а,b and 39 (Hodge \emph{et al.}~1990)}

Valdez-Gutierrez \emph{et al.}~(2001) provided the following parameters averaged over
the group of objects nos.~40а,b and~39: the velocity at the line peak,
${-239}$~km~s$^{-1}$, and the velocity dispersion, 10.4~km~s$^{-1}$. We separately
scanned the two bright sources no.~40a (scan~16) and no.~40b (scan~15); both also
include the emission from the more extended and fainter region no.~39. As follows from
Fig.~3f, the velocity at the peak of these three regions is ${-240}$~km~s$^{-1}$. The
H~II region no.~39 is characterized by a narrow H$\alpha$ line with no wings with the
FWHM determined by the IFP501 instrumental profile along the entire length. At the same
time, both bright compact H~II regions nos.~40a and~40b (scans~16 and~15, respectively)
exhibit weak emission at $10\%$ $I(\max)$ in the velocity range from ${-310}$ to
${-175}$~km~s$^{-1}$.

\subsection*{The Supernova Remnant}

The bright nebula~S8 (Sandage~1971), the supernova remnant, shows
emission over the entire velocity range from 0 to
${-600}$~km~s$^{-1}$ determined by the IFP501 free spectral range
(scans~15 and~16, 19\mbox{--}23~arcsec). The remnant emission is
observed with IFP235 at least in the velocity range from ${+200}$
to ${-600}$~km~s$^{-1}$ (scan~17). An interfering emission
emerges at higher velocities, ${+400}$ and ${-800}$~km~s$^{-1}$.
This emission is probably attributable to an improper allowance
for the night-sky 6554~\AA\, line.

This result is in complete agreement with previous observations of Lozinskaya \emph{et
al.}~(1998) and Rosado \emph{et al.}~(2001).

\begin{table}[t!]
\caption{Parameters of the bright ionized shells in the complex}
\begin{tabular}{c|c|c|c|l}
\hline Shell& \parbox[c][1cm]{1.5cm}{ Size, pc} &
\parbox[c][1cm]{1cm}{ $V_{{exp}}$, km~s$^{-1}$} &
\parbox[c][1cm]{2cm}{$V_{{exp}}$(V--G), km~s$^{-1}$}&
\multicolumn{1}{c}{\parbox[c][1cm]{1.5cm}{Age, Myr }}\\
\hline
R1      &  $188\times138$    & 60--75    &  29         & \,0.7 (81/67)  \\
R2      &  $145\times88$\phantom{9}& 50--60    &  49         & \,0.6 (58/55) \\
R3      &  $258\times209$    &            &  30         & \,1.9 (V--G)    \\
R4      &  $234\times138$    & 40--45    &  32         & \,2.2 (93/42)   \\
R5      &  $113\times113$    & 30--50    &  38         & \,0.8 (56/40)  \\
R6      &  $226\times184$    & $\leq 20$   &  25         & \,2.0 (V--G)     \\
R8      &  $217\times154$    &$\simeq 65$ &  26  & \,1.7 (185/65)  \\
\hline
\end{tabular}
\end{table}

\bigskip

All of the ionized shells in the complex of star formation for which we managed to
reliably detect their expansion based on the velocity ellipse are collected in Table~3.
For comparison, the fourth column of the table gives the expansion velocities of the same
shells determined by Valdez-Gutierrez \emph{et al.}~(2001) from line-profile splitting.
Since the shell sizes were measured from our deep images, they slightly differ from
those of Valdez-Gutierrez \emph{et al.}~(2001). It should be noted that, given the
irregular shape of most shells, the representation of the observed shells by regular
ellipses is arbitrary and subjective in nature.

In estimating the kinematic ages of the shells from their radii
and expansion velocities, we used the classical theory of  wind
blown bubble ($t=0.6~R/V$). In addition to the age, the mean
radius and expansion velocity used for its estimation are given
in parentheses (in the form~$R/V$) in the last column of the
table. The ages of shells~R3 and~R6 correspond to the data from
Valdez-Gutierrez \emph{et al.}~(2001).

\section{DISCUSSION}

The radial-velocity variations with distance from the center of the ionized shells
(velocity ellipses) found here are generally in agreement with the observations of
Meaburn \emph{et al.}~(1988) and Valdez-Gutierrez \emph{et al.}~(2001) but they give a
clearer picture of expansion. In the latter paper, the conclusions regarding expansion
were drawn from integrated line profile splitting in the shells. Since bright peripheral
regions or individual bright clumps can give a significant contribution when averaging
over the shell or over its separate fields, the expansion velocity determined in this
way can be underestimated. The large brightness difference between the approaching and
receding sides of the shells can also cause the expansion velocity inferred from line
splitting to be underestimated. The constructed velocity ellipses allow these
difficulties to be circumvented. The expansion velocity can be determined from the
velocity ellipse by taking into account the geometrical projection even if only one
shell side is observed, as in the case of~R8. Indeed, as we see from Table~3, our
estimates of the expansion velocity in most shells gave higher values than those in
Valdez-Gutierrez \emph{et al.}~(2001).

In several shells, we detected a distinct asymmetry in their expansion: the approaching
and receding sides have different velocities.

On scans~v1 and~v2 crossing the entire chain of the brightest
ionized shells, we clearly see that all of them generally have
similar velocities but an irregular, clumpy brightness
distribution. This irregular brightness distribution may have
served as the basis for the conclusion of Valdez-Gutierrez
\emph{et al.}~(2001) that shells~R1 and~R2 are at different
distances, because they are observed at different radial
velocities. These scans show that individual bright clumps on the
two sides of these shells actually have different velocities but
they all fall within our velocity ellipses. According to our
measurements, the approaching and receding sides of the two
shells have similar velocities: $-290\div-320$ and
$-170\div-190$~km~s$^{-1}$, as inferred from different scans
in~R1; $-280\div-300$ and ${-170}$~km~s$^{-1}$ in~R2. The far
side is brighter in~R1 and the near side is brighter in~R2. We
also noted that the southern part of shell~R1 recedes faster (${V
= -150}$~km~s$^{-1}$) than does its northern part (${V =
-180}$~km~s$^{-1}$).

Taking into account the structure of~R1 and~R2 in different velocity ranges, we do not
rule out the possibility that these shells generally constitute a single dumbbell-like
structure formed on both sides of a dense neutral-gas layer. However, this assumption
requires additional observational confirmation.

Valdez-Gutierrez \emph{et al.}.~(2001) showed that the  energy of
the stellar wind from nearby ОВ~associations was enough for most
of the ionized shells to be formed. The rapidly expanding
shells~R1 and~R2 constitute an exception. We further increased the
expansion velocity of these two objects (see Table.~3) but
simultaneously found additional, most intense sources of the
stellar wind at their boundary --- early-type supergiants and
giants (Lozinskaya \emph{et al.}~2002), which removes the problem
of the mechanical energy sources.

Recall also that the two mentioned shells~R1 and~R2 exhibit an
intense emission  in the [S~II] 6717/6731~\AA\, lines (see
Valdez-Gutierrez \emph{et al.}~2001; Lozinskaya \emph{et
al.}~2002), typical of the radiation of the shock waves generated
by a supernova explosion.

The neutral shells produced by the combined effect of the wind and supernovae in
ОВ~associations are being widely searched for in nearby galaxies (see, e.g., the review
article by Brinks~1994; Kim \emph{et al.}~1999; Oey \emph{et al.}~2002; and references
therein). In most cases, the 21-cm observations either do not reveal any distinct
neutral-gas shell structure around ОВ~associations at all or do not allow the
neutral-gas structures to be unambiguously associated with H~II shells. The studies of
neutral gas in the vicinity of the three well-known ionized supershells in the LMC
carried out by Oey \emph{et al.}~(2002) serve as a clear example.

The complex of star formation in IC~1613 constitutes a lucky exception. We do not know
any other examples of such a distinct interaction between multiple ionized and neutral
shells that is observed in the complex under discussion.

Here, we have studied the kinematics of the identified neutral shells~I, II, and~III for
the first time. We have also detected expansion and measured the expansion velocity for
the first time.

As was mentioned in Section~3, the derived radii and expansion velocities of the H~I
shells in IC~1613 fall within the region of the peak in the distribution of neutral
shells in the LMC and SMC in size and expansion velocity. The age of the H~I shells in the
complex of star formation in IC~1613 (5.3\mbox{--}5.5~Myr) is also in complete agreement
with the peaks in the age distributions of H~I supershells in the LMC (4.9~Myr; Kim
\emph{et al.}~1999) and in the SMC (5.4~Myr; Staveley-Smith \emph{et al.}~(1997).

We estimated the kinematic ages of the neutral shells~II and~III in Table~2 from their
mean radii and expansion velocities by using the classical theory of a constant stellar
wind in a homogeneous medium. This assumes that the mechanical luminosity of the wind
from the association stars responsible for the shell formation does not vary with time
and that all stars were formed simultaneously. Of course, both assumptions are not valid,
although they are universally accepted. Allowance for sequential star formation and for
variations in the mass loss rate during the evolution of a rich association as well as
for the cloudy structure of the interstellar medium can change the age estimate by a
factor of~1.5 to~3 (see, e.g., Shull and Saken~1995; Oey \emph{et al.}~1996; Silich
\emph{et al.}~1996; Silich and Franco~1999; and references therein). Nevertheless, we
may conclude that the derived ages of shells~II (5.6~Myr) and~III (5.3~Myr) agree with
the ages of the stellar associations in the complex of star formation. Shell~I, whose
expansion we failed to detect, is similar in size to shells~II and~III and is most
likely similar in age.

The three-color photometry of stars in IC~1613 performed by Hodge \emph{et al.}~(1991)
gave a minimum age of the stars in their field no.~1, where the neutral shells that we
identified are located, equal to 5~Myr. This field includes associations nos.~10, 12,
13, 14, 15, and~17 from the list by Hodge~(1978), whose mean and minimum ages were
estimated by Hodge \emph{et al.}~(1991) to be 17 and 3~Myr, respectively. Recent
estimates by Georgiev \emph{et al.}~(1999) yielded similar results: the age of the
youngest nearby associations nos.~10 and~14 is about 5~Myr and the age of the oldest
associations (nos.~12,~19, and~19) reaches 20~Myr.

The mechanical wind luminosity that is required for the neutral
shells observed in the complex to be formed and that is
determined by using the classical model (with all of the
reservations made above) is $5\times10^{38}$, $5\times10^{37}$,
and $5\times10^{36}$~erg~s$^{-1}$ in a medium with an initial
density of 10, 1, and 0.1~cm$^{-3}$, respectively. These values
seem reasonable enough for the stars found in the region --- the
sources of a strong stellar wind (see Valdez-Gutierrez \emph{et
al.}~2001; Lozinskaya \emph{et al.}~2002). It should be
emphasized, however, that the spectra of bright stars in the
region are required to adequately discuss the sources of
mechanical energy. Lozinskaya \emph{et al.}~(2002) obtained the
spectra of stars only in three small fields of the star-forming
complex and detected stars with strong winds, blue supergiants
and Of~star, in two of them.

Despite the ``suitable'' age and input of mechanical energy, we do not consider it
possible to unequivocally associate the formation of neutral shells with the action of
the stellar wind from the OB~associations identified by Georgiev \emph{et al.}~(1999) for
the following reasons.

First, most of the H~I supershells identified in the LMC, the
SMC, and other Local-Group galaxies are several-fold younger than
the corresponding OB~associations (see, e.g., Kim \emph{et
al.}~1999; Staveley-Smith \emph{et al.}~1997, and references
therein). Therefore, it may well be that the H~I shells that we
detected in IC~1613 were also formed by an older population than
the young groups of stars shown in Fig.~1c.

As follows from Table~3 and from similar estimates by Valdez-Gutierrez \emph{et
al.}~(2001), the ages of most of the ionized shells in the complex lie within the range
from 0.6 to 2.2~Myr. These ages are much younger than the ages of the OB~associations
that could be responsible for their formation. Such a situation is observed in most
galactic and extragalactic ionized shells and supershells. In general, to explain this
mismatch, either noncoeval star formation is assumed or the shells are assumed to be
formed only by stars at a late evolutionary stage --- WR, Of, and BSG with a
short-duration, but intense wind. The situation in the complex under consideration is
more complicated, because the bright ionized shells are localized inside or at the
boundary of the older neutral shells and their evolution differs significantly from the
standard theory.

In addition, a number of observed facts actually suggest noncoeval star formation in the
complex.


Figure~1c suggests that all of the young OB~associations in the region are localized not
at the center but in the dense peripheral parts of the H~I shells [the only association
no.~7 from the list by Georgiev \emph{et al.}~(1999) is partly located inside
shell~III). In general, this picture may provide evidence for the formation of these
associations in dense H~I shells. We emphasize that the new association boundaries
delineated by Georgiev \emph{et al.}~(1999) split the associations of Hodge~(1978) into
smaller groups, which are believed to be the young nuclei of the corresponding
associations. The scenario for triggered stars formation in expanding shells was
considered by many authors [see, e.g., Elmegreen \emph{et al.}~(2002) and references
therein].

Comparison of the ages and mutual localizations of the ionized and neutral structures in
the complex also suggests noncoeval star formation. All the ionized shells are
several-fold younger than the neutral shells. As was pointed out in Section~3, most of
the bright ionized shells are located in the dense peripheral parts of the H~I shells.
The only exception is shell~R4, which completely fits into the neutral shell~II from the
inside. According to our measurements, this ionized shell is oldest and all the young
nuclei of the OB~associations are located in its boundary regions (see Fig.~1c).

However, these arguments for sequential or triggered star formation, which refer to the
neutral shells of the complex studied in detail here, are speculative, because the age
difference between the H~I and H~II shells is small.

The multishell complex itself, which represents the only region
of violent ongoing star formation in IC~1613, may have been
produced by the collision of two older and more massive giant
neutral supershells (Lozinskaya~2002a, 2000b). Indeed, in
addition to the three bright and relatively small H~I shells
considered here, a giant H~I ring south of the star-forming
complex and a giant arc structure in the north, which probably
also represents part of the neutral supershell, can be identified
in Fig.~2a. The characteristic size of the two structures is
${\sim1{-}1.5}$~kpc. The complex of ongoing star formation lies
at their common boundary, where the collision of these two giant
supershells could trigger violent star formation. We are planning
to consider this scenario in detail.

\section{CONCLUSIONS}

We studied in detail the structure and kinematics of the neutral and ionized gas
components in the only known complex of star formation in the irregular dwarf galaxy
IC~1613.

To study the kinematics of the ionized shells, we carried out H$\alpha$ observations with
a scanning Fabry--Perot interferometer attached to the 6-m SAO telescope. The
monochromatic H$\alpha$ image of the multishell complex obtained from our
interferometric observations reveals new faint filamentary structures in several regions
of the complex.

We constructed position\mbox{--}radial velocity diagrams, which cover the entire complex
of star formation almost uniformly. The characteristic velocity variation with distance
from the center, the velocity ellipse, was used to refine (increase) the expansion
velocities of most ionized shells in the complex estimated by Valdez-Gutierrez \emph{et
al.}~(2001). The expansion in several shells was found to be asymmetric: the approaching
and receding sides of the shells have different velocities.

Based on our VLA 21-cm observations, we have studied the neutral-gas kinematics in the
complex of star formation for the first time. The mean H~I velocity in the complex is
$ХV_{\textrm{Hel}} = -230\pm 5$~km~s$^{-1}$, in good agreement with the estimate
obtained by Lake and Skillman~(1989) for this part of the galaxy from
low-angular-resolution observations.

We identified three extended (300\mbox{--}350~pc) neutral shells with which the
brightest ionized shells in the complex of star formation are associated. The two H~I
shells were found to expand at a velocity of 15\mbox{--}18~km~s$^{-1}$.

The sizes, expansion velocities, and kinematic ages of the neutral shells in the complex
fall within the regions of the peaks in the corresponding distributions for giant shells
in the LMC and SMC.

We identified an incomplete H~I shell with the only known supernova remnant in the
galaxy located at its inner boundary. This confirms the scenario for a supernova
explosion inside a cavity surrounded by a dense shell and a collision of the remnant
with the shell wall suggested by Lozinskaya \emph{et al.}~(1998) to explain the
peculiarity of this remnant, which combines the properties of young and old objects.

We found evidence of the physical interaction between the H~I and H~II shells in the
region of the chain of stars, early-type giants and supergiants, detected by Lozinskaya
\emph{et al.}~(2002). The region at the boundary of the two shells where the stellar
chain is located was shown to be the dynamically most active part of the star-forming
complex. The highest H~II and H~I velocities are observed here.

The relative positions and ages of the H~I and H~II shells and OB~associations in the
complex suggest sequential or triggered star formation in the expanding neutral shells.

In addition to the three brightest and most prominent H~I shells, we found supergiant
arches and ring structures in the galaxy whose sizes are comparable to the gaseous-disk
thickness. These may be assumed to be the traces of preceding starbursts in IC~1613.

\section*{ACKNOWLEDGMENTS}

This study was supported by the Russian Foundation for Basic Research (project
nos.~01-02-16118 and 02-02-06048mas), the Federal Central Research Program (contract
40.022.1.1.1102), and CONACYT (Mexico, project 36132-Е). The observational data were
obtained with the 6-m SAO telescope financed by the Ministry of Science of Russia
(registration number 01-43). We are grateful to the 6-m Telescope Committee for
allocating observational time. The National Radio AStronomy Observatory (NRAO) belongs
to the National Science Foundation (USA) and is operated by the Association of
Universities Inc. under a contract with the NSF. We wish to thank S.A.~Silich who read
the manuscript for helpful remarks.

Translated by V. Astakhov

\end{document}